\documentclass{article}
\begin{document}
\title{Analysing Relations involving small number of Monomials in AES S- Box}
\author{Riddhi Ghosal \\ Indian Statistical Institute \\ Email: postboxriddhi@gmail.com}
\maketitle
\begin{abstract}
In the present day, AES is one the most widely used and most secure Encryption Systems prevailing. So, naturally lots of research work is going on to mount a significant attack on AES. Many different forms of Linear and differential cryptanalysis have been performed on AES. Of late, an active area of research has been Algebraic Cryptanalysis of AES, where although fast progress is being made, there are still numerous scopes for research and improvement. One of the major reasons behind this being that algebraic cryptanalysis mainly depends on I/O relations of the AES S- Box (a major component of the AES). As, already known, that the key recovery algorithm of AES can be broken down as an MQ problem which is itself considered hard. Solving these equations depends on our ability reduce them into linear forms which are easily solvable under our current computational prowess. The lower the degree of these equations, the easier it is for us to linearlize hence the attack complexity reduces. The aim of this paper is to analyze the various relations involving small number of monomials of the AES S- Box and to answer the question whether it is actually possible to have such monomial equations for the S- Box if we restrict the degree of the monomials. In other words this paper aims to study such equations and see if they can be applicable for AES.
\end{abstract}
\section{Introduction}
In [2] there is a detailed description of writing IO equations and monomial relations for the 6*4 DES S- Box. In this paper I have tried to emulate a similar kind of approach for the AES S- box as well. According to the authors of the above mentioned document, they encountered few equations that either contained a very few number of monomials or they were multivariate equations of low degree. 
The idea of monomials shall be discussed here for the 8*8 AES S- box. The main purpose of this paper is to see if the monomials discussed in [2] can be implemented for AES as well. [2] talks about relations involving at maximum 4 monomials. Their aim was to keep the sum of degrees of all monomials in an equation less than or equal to 15. Here I deal with 5 monomials as well and the sum of degrees of monomials can be 25 at maximum. My aim is to check if such types of equations are actually possible if we restrict their degrees by some upper bound. The conclusion drawn after executing several programs (which can be obtained from the author) was that we do not get any such monomial equations if we restrict the degree to 5. For equations of degree greater than 5, it may become possible to make such constructions.

\section{Notations}
Let the input bits for the S- box be X1, X2, X3, X4, X5, X6, X7, X8. \\
Let the output bits for the S- box be Y1, Y2, Y3, Y4, Y5, Y6, Y7, Y8. \\
There are 256 possible input states for the S- box, let these states be named as S0, S1, S2, ….. , S254, S255.\\
a, b, c, d, e are symbols used to denote monomial variables.

\section{Analysis of Equations}
Before further analysis I would like to bring to notice one important observation, which is one of the main reasons why these kinds of relations are not possible in an AES S- Box. \\
We see that S8, S9, S48, S64, S82 have the least number of the expressions of the form a - 1 = 0. (7 such expressions to be precise) \\ \\  In this table X5 implies X5 - 1 =0 and so on.

\begin{table}[ht]
\caption{Input States with least numberof variables equal to 1}
\centering
\begin{tabular}{c c c c c c c c}
\hline\hline
Input State \\ [0.5ex]
\hline
S8 & X5 & Y3 & Y4 & X5Y3 & X5Y4 & Y3Y4 & X5Y3Y4\\
S9 &	X5 &	X8 &	Y8 &	X5X8 &	X5Y8	& X8Y8 &	X5X8Y8\\
S48	& X3	& X4	& Y6	& X3X4	& X3Y6	& X4Y6	& X3X4Y6\\
S64	& X2	& Y5	& Y8	& X2Y5	& X2Y8	& Y5Y8	&X2Y5Y8\\
S82	& X2	 & X4 &	X7	& X2X4	& X2X7	& X4X7	& X2X4X7\\ [1ex]
\hline
\end{tabular}
\label{table:1}
\end{table}

\subsection{Equations with 1 monomial}
First let us analyse the relations involving 1 monomial. By monomial we mean equations of type X1X2Y2 = 0 (say). \\
For an AES S- box, it can be easily shown that the number of such monomial relations are 20774. 
But when we restrict the degree of these monomials to 5, the number is surprisingly 0. Hence we have no monomial equations of very low degree. The algorithm used to check the number of monomials is available with the author and can be obtained on request.

\subsection{Equations with 2 monomials}
When we say 2 monomials or binomials, we mainly deal with 2 cases:\\
a)	Equations of the form say a + b = 0 \\
b)	Equations of the form a + 1 = 0\\

We analyze the latter first. In order for an expression of the form a + 1 = 0, we must have a = 1 for all the possible input states. Table 1 clearly shows that no such binomial exists for which a - 1 = 0 for all input states when the degree is restricted to 5 (in fact this is impossible even if we consider binomials of all possible degrees).
\\
Now for the former case, if a + b = 0 then a = b = 0 or a = b = 1 for all possible input states.
a = b = 0 is trivially not possible as we have seen clearly that no such monomials exist for which a = 0 for all inputs. Similarly, the case where a = b = 1 is also ruled out from table 1. 

\subsection{Equations with 3 monomials}
 Again this can subdivided into two cases:\\
a)	Equations of the form a + b + 1 = 0\\
b)	Equations of the form a + b + c = 0\\

For first case either of a or b must be equal to 1. In other word a and b should always have alternating values. In case both have the same value then the relation is not satisfied. A practically implemented program (code available with author) has shown that no such a and b exist such that they have alternate values for all input states.\\
In the second case a = b = c = 0 that is not possible as number of monomials which are equal to 0 for all inputs are 0. An alternative is any two of a, b, c = 1 and the third one is 0. Let us say that a = b =1. Now from table 1, it is clear that there cannot exist any unique a and b that will satisfy the above expression. 

\subsection{Equations with 4  monomials}
a)	 Equations of the form a + b + c + d = 0\\
b)	Equations of the form  a + b + c + 1 = 0\\
For the first case there are three alternatives, all four 1’s, two 1’s and two 0’s, all four 0’s. The first and the last options can be ignored. So we analyze the only remaining choice. \\
We need two 1’s, so for satisfying S8, S9, S48, S64 and S82, from the given monomials we must choose exactly two from each of these states. A close observation will show that the best possible choice can be X4 + X2 + Y8 + X5. But this does not satisfy S8 and S48. No other possible choice of variables can make this relation true.\\
For case b) we again have two choices all three a, b, c =1 or any 1 of a, b, c =1 and the other two should be 0. From table 1 we again have that first choice is impossible. A simple algorithm correctly implemented in a program reveals that such a combination is not possible either. A closer look at table 1 shows that if we need an expression that satisfies S8, S9, S48, S64 and S82, the only choice we have is X2 + (X4)(X3)(Y6) + X5 + 1 = 0. ( () implies either one) Unfortunately S1 itself does not satisfy this expression.

\subsection{Equations with 5  monomials}
a)	Equations of the form a + b + c + d + e =  0\\
b)	Equations of the form a + b + c + d +  1 = 0\\
For the former the possible alternatives are four 1’s one 0, two 1’s three 0’s, all 0’s. First and last option can be trivially ignored. The second option needs a closer observation of Table 1 again. The best possible option here can be X4 + X2 + Y8 + X5 + (?). I prefer to leave a question mark here because regardless of what your choice is, either one of S9 or S49 will always be dissatisfied.\\ 
Now let us see the second case. Three 1’s one 0, one 1 three 0’s are the only possibilities. First choice is not applicable from Table 1.  Well for the second choice we can check all possible ways so that S8, S9, S48, S64 and S82 fit in, but all such expressions will be defied by S255. 

\section{Conclusion}
Thus we see that equation of low degree containing small number monomials is not really possible for an AES S- Box. This is one of the reasons because of which mounting an effective Algebraic Attack on AES becomes difficult.\\
If we allow monomials more than degree 5 then such combinations are definitely possible. Number of monomials relations that exist when degree is 6 is 43. Also if the number of monomials are allowed more than 5 then perhaps such constructions may be possible, but these expressions tend to get quite large hence maybe difficult to handle.

\section{Acknowledgements}
I would like give my sincere thanks to Dr. Goutam Paul of Indian Statistical Institute who had introduced this topic to me as a part of my summer project under his supervision.

\end{document}